\documentclass[pre,superscriptaddress]{revtex4}
\usepackage{amssymb}
\usepackage{amsmath}
\usepackage{color}
\usepackage{graphicx}
\usepackage{mathrsfs}

\newcommand{\pa}{\partial}

\begin{document}
\baselineskip 16pt
\parindent=2em
\title{
Solutions For A Generalized Fractional Anomalous Diffusion Equation}
\author{Long-Jin Lv}
\email{Lojin@stu.hdu.edu.cn} \affiliation {School of Science,
Hangzhou Dianzi University, Hangzhou 310037, China }

\author{Jian-Bin Xiao}
 \affiliation {School of Science, Hangzhou Dianzi University, Hangzhou 310037, China }

\author{Lin Zhang}
 \affiliation {School of Science, Hangzhou Dianzi University, Hangzhou 310037, China }
\author{Lei Gao}
\affiliation {School of Science, Hangzhou Dianzi University,
Hangzhou 310037, China }

\begin{abstract}
In this paper, we investigate the solutions for a generalized
fractional diffusion equation that extends some known diffusion
equations by taking a spatial time-dependent diffusion coefficient
and an external force into account, which subjects to the natural
boundaries and the generic initial condition. We obtain explicit
analytical expressions for the probability distribution and study
the relation between our solutions and those obtained within the
maximum entropy principle by using the Tsallis entropy.
\end{abstract}
\maketitle
 \hspace{8mm} {\bf Keywords:}\quad Anomalous diffusion; Fractional diffusion;
 Green function; Fox function

\section{Introduction}

Recently, anomalous diffusion equations have been extensively
investigated due to the broadness of their physical applications. In
fact, fractional diffusion equations and the non-linear fractional
diffusion equations have been successfully applied to several
physical situations such as percolation of gases through porous
media [1], thin saturated regions in porous media [2], in the
transport of fluid in porous media and in viscous fingering[3], thin
liquid films spreading under gravity [4], modeling of non-Markvian
dynamical processes in protein folding [5], relaxation to
equilibrium in system (such as polymer chains and membranes) with
long temporal memory [6], anomalous transport in disordered systems
[7], diffusion on fractals [8], and the multi-physical transport in
porous media, such as electroosmosis[9-10]. Note that the physical
systems mentioned above essentially concern anomalous diffusion of
the correlated type (both sub and super-diffusion; see [11] and
references therein) or of the L\'{e}vy type (see [12] and references
therein). The anomalous correlated diffusion usually has a finite
second moment $\langle x^2\rangle\propto t^\sigma$
($\sigma>1,\sigma=1$ and $0<\sigma<1$ correspond to super-diffusion,
normal diffusion and sub-diffusion, respectively; $\sigma=0$
corresponds basically to localization). Due to the broadness of the
problems involving anomalous diffusion, one needs to apply different
kinds of theoretical approaches such as nonlinear Fokker-Planck
equation (or modified porous media equation), fractional
Fokker-Planck equation, Fokker-Planck equation with spatial
dependent diffusion coefficient , and generalized Langevin
equations. The properties concerning these equations have been
intensively investigated [13-17] and the lattice Boltzmann method
was used to get the numerical solutions for this equations which
govern the multi-physical transfort in porous media.

In order to cover the above situations, we employ a spatial
time-dependent diffusion coefficient, in other words, our work is
aimed at the investigation of solutions for a fractional diffusion
equation taking a spatial time-dependence on the diffusion
coefficient and an external force (drift) into account. More
precisely,we focus our attention on the following equation:
\begin{equation}\label{1}
\frac{\pa^{\gamma}}{\pa
t^{\gamma}}\rho(x,t)=\int_{0}^{t}dt'\frac{\pa}{\pa
x}\{D(x,t-t')\frac{\pa^{\mu-1}}{\pa
x^{\mu-1}}[\rho(x,t)]^{\nu}\}-\frac{\pa}{\pa x}\{F(x)\rho(x,t)\},
\end{equation}
where $0 <\gamma \leq1,\mu,\nu\in R$, the diffusion coefficient is
given by $D(x,t)=D(t)|x|^{-\theta}$, which is a spatial
time-dependent diffusion coefficient,and $F(x)=-\frac{\pa V(x)}{\pa
x}$is an external force (drift) associated with the potential
$V(x)$. Here, we use the Caputo operator [18] for the fractional
derivative with respect to time $t$, and we work with the positive
spatial variable $x$. Later on, we will extend the results to the
entire real $x$-axis by the use of symmetry (in other words, we are
working with ${\pa}/{\pa|x|}$ and ${\pa^{\mu -1}}/{\pa|x|^{\mu
-1}}$). Also, we employ, in general, the initial condition
$\rho(x,0)=\tilde{\rho}(x)$ ($\tilde{\rho}(x)$ is a given function),
and the boundary condition $\rho(x\!\to\!\pm\infty,t)\rightarrow 0$.
For Eq.(1), one can prove that $\int_{-\infty}^{+\infty}dx\rho(x,t)$
is time independent (hence, if $\rho (x,t)$ is normalized at
$t=0$,it will remain so forever). Indeed, if we write Eq.(1) as
$\pa_{t}^{\gamma}=-\pa_{x}J$, and, for simplicity, assume the
boundary conditions $J(\pm\infty)=0$, it can be shown that
$\int_{-\infty}^{+\infty}dx\rho(x,t)$ is a constant of motion (see
[19] and references therein). Note that when
$(\mu,\gamma,\nu)=(2,1,1)$, Eq.(1) recovers the standard
Fokker-Planck equation in the presence of a drift taking memory
effects into account. The particular case $F(x)=0$ (no drift) and
$D(x,t)=D\delta(t)$ with $(\mu,\gamma)=(2,1)$ has been considered by
spohn [20]. The case $D(x,t)=D\delta(t)|x|^{-\theta}$ with
$(\mu,\nu)=(2,1)$ and the case $D(x,t)=D(t)$ with $(\mu,\nu)=(2,1)$
have been investigated in [21] and [22], respectively.

Explicit solutions play an important role in analyzing physical
situations, since they contain, in principle, precise information
about the system. In particular, they can be used as an useful guide
to control the accuracy of numerical solutions. For these reasons,
we dedicated to this work to investigate the solutions for Eq.(1) in
some particular situations. In all the particular cases, Eq.(1)
satisfies the initial condition $\rho(x,0)=\tilde{\rho}(x)$
($\tilde{\rho}(x)$ is a given function), and the boundary condition
$\rho(\pm \infty,t)=0$. The remainder of this paper goes as follow.
In Sec.2, we obtain the exact solutions for the special cases. In
Sec.3, we present our conclusions.

\section{ Exact solutions for different case}

In this section, we start our discussion by considering  Eq.(1) in
the absence of external force with
$D(t)=Dt^{\alpha-1}/\Gamma(\alpha)$ (
$D(t)=D\delta(t)$),$(\mu,\nu)=(2,1)$ and $\gamma$, $\theta$
arbitrary. For this case, Eq.(1) reads
\begin{equation}\label{2}
\frac{\pa^{\gamma}}{\pa
t^{\gamma}}\rho(x,t)=\int_{0}^{t}dt'D(t-t')\frac{\pa}{\pa
x}\{|x|^{-\theta}\frac{\pa}{\pa x}{\rho(x,t)}\}.
\end{equation}
Here, we use the Caputo operator [18] for the fractional derivative
with respect to time $t$. By employing the Laplace transform in
Eq.(2), we obtain
\begin{equation}\label{3}
\tilde{D}(s)\frac{\pa}{\pa x}\{|x|^{-\theta}\frac{\pa}{\pa
x}\tilde{\rho}(x,s)\}-s^{\gamma}\tilde{\rho}(x,s)=-s^{\gamma-1}\rho(x,0),
\end{equation}
where $\tilde{\rho}(x,s)=\mathscr{L} \{\rho(x,t)\}$,
$\tilde{D}(s)=\mathscr{L}\{D(t)\}$, and $
\mathscr{L}\{f(t)\}=\int_0^\infty dte^{-st}f(t)$ denotes the Laplace
transform of the function $f$. This equation can be solved by Green
function method [23]. By substituting
\begin{equation}\label{4}
\tilde{\rho}(x,s)=\int dx'\tilde{\mathcal
{G}}(x-x',s)\tilde{\rho}(x'),
\end{equation}
into Eq.(3) which yields
\begin{equation}\label{5}
\tilde{D}(s)\frac{\pa}{\pa x}\{|x|^{-\theta}\frac{\pa}{\pa
x}\tilde{\mathcal {G}}(x,s)\}-s^{\gamma}\tilde{\mathcal
{G}}(x,s)=-s^{\gamma-1}\mathcal {G}(x,0).
\end{equation}
where $\mathcal {G}(x,t)$ subjects to the initial condition
$\mathcal {G}(x,0)=\delta(x)$ and the boundary condition $\mathcal
{G}(\pm\infty,t)=0$.

In order to solve Eq.(5), it is convenient to perform the transform
[24]
\begin{equation}\label{6}
y=A(s)x^{v},\quad \mathcal {G}(x,s)=y^{\delta}Z(y)
\end{equation}
to translate Eq.(5) into the second-order Bessel equation as
\begin{equation}\label{7}
y^2\frac{\pa^2 Z}{\pa y^2}+y\frac{\pa Z}{\pa
y}-(\lambda^2+y^2)Z(y)=-\frac{y^{2-\delta}}{s}\delta((\frac{y}{A(s)})^{\frac{1}{v}})
\end{equation}
with parameter $\lambda^2$ under the following conditions:
\begin{equation}\label{8}
v=\frac{2+\theta}{2},A(s)=\frac{1}{v}[\frac{s^\gamma}{\tilde{D}(s)}]^\frac{1}{2},\lambda=\frac{1+\theta}{2+\theta},\delta=\frac{1+\theta}{2+\theta}.
\end{equation}
Since Eq.(5) should fit the boundary condition $\mathcal
{G}(\pm\infty,t)=0$, i.e. $\mathcal {G}(\pm\infty,s)=0$, we get the
solution of Eq.(5)
\begin{equation}\label{9}
\mathcal {G}(x,s)=C(s)y^{\delta}K_{\lambda}(y),
\end{equation}
where $K_{n}(x)$ is the modified Bessel function of second kind; and
$C(s)$ can be determined by the normalization of $\mathcal
{G}(x,t)$, i.e. $\int_0^\infty dx\tilde{\mathcal
{G}}(x,s)=\frac{1}{2s}$. After some calculations, we obtain
\begin{equation}\label{10}
\tilde{\mathcal
{G}}(x,s)=\frac{2+\theta}{\Gamma(\frac{1}{2+\theta})s}
(\frac{1}{2+\theta}(\frac{s^{\gamma}}{\tilde{D}(s)})^\frac{1}{2})^{\frac{3+\theta}{2+\theta}}|x|^{\frac{1+\theta}{2}}
K_{\frac{1+\theta}{2+\theta}}
(\frac{2}{2+\theta}(\frac{s^{\gamma}}{\tilde{D}(s)})^\frac{1}{2}|x|^{\frac{2+\theta}{2}}),
\end{equation}
where, we used the formula
\begin{equation}\label{11}
\int_0^\infty dy\cdot y^v
K_\lambda(ay)=2^{v-1}a^{-v-1}\Gamma(\frac{1+v+\lambda}{2})\Gamma(\frac{1+v-\lambda}{2}).
\end{equation}
\begin{raggedleft}
\textbf{  Case 1. } $D(t)=D\delta(t)$, i.e. $\tilde{D}(s)=D$.
\end{raggedleft}

Since $K_\lambda (x)=\frac{1}{2}H_{0 \quad 2}^{2 \quad
0}[\frac{x^2}{4}|_{(-\lambda /2,1)(\lambda /2,1)}]$, we can get the
Laplace inverse of $\tilde{\mathcal {G}}(x,s)$ by applying the
property of the Laplace inverse of Fox function, which yields
\begin{equation}
\mathcal
{G}(x,t)=\frac{2+\theta}{2\Gamma(\frac{1}{2+\theta})}(\frac{1}{(2+\theta)^2
Dt^{\gamma}})^{\frac{1}{2+\theta}}H_{1 \quad 2}^{2\quad
0}[\frac{|x|^{2+\theta}}{(2+\theta)^2
Dt^{\gamma}}|_{(0,1),(\frac{1+\theta}{2+\theta},1)}^{(1-\frac{\gamma}{2+\theta},\gamma)}],
\end{equation}
where $H_{p\quad q}^{m\quad
n}[x|_{(b_1,B_1),...,(b_q,B_q)}^{(a_1,A_1),...,(a_p,A_p)}]$ is the
FOX function [25]. Thus, we can find the solution by substituting
Eq.(10) into Eq.(4), which yields
\begin{equation}
\rho(x,t)=\frac{2+\theta}{2\Gamma(\frac{1}{2+\theta})}(\frac{1}{(2+\theta)^2
Dt^{\gamma}})^{\frac{1}{2+\theta}}\int_{-\infty}^{+\infty}dx'\tilde{\rho}(x')H_{1
\quad 2}^{2\quad 0}[\frac{|x-x'|^{2+\theta}}{(2+\theta)^2
Dt^{\gamma}}|_{(0,1),(\frac{1+\theta}{2+\theta},1)}^{(1-\frac{\gamma}{2+\theta},\gamma)}].
\end{equation}
In fig.1, we show the behavior of the above equation by considering
typical values of $\gamma$ and $\theta$ with
$\tilde{\rho}(x)=\delta(x)$. At this point, it is interesting to
analyze the asymptotic behavior of Eq.(13). For simplicity, we
consider $\tilde{\rho}(x)=\delta(x)$, so
$\rho(x,t)=\mathcal{G}(x,t)$; and the asymptotic behavior of
$\rho(x,t)$ is
$$\rho(x,t)\sim
\frac{2+\theta}{2\Gamma(1/(2+\theta))}(2-\gamma)^{-\frac{1}{2}}\gamma^{\frac{\gamma}{(2+\theta)(2-\gamma)}-\frac{1}{2}}(\frac{1}{(2+\theta)^2
Dt^{\gamma}})^{\frac{1}{(2+\theta)(2-\gamma)}}|x|^{\frac{\gamma-1}{2-\gamma}}$$
\begin{equation}
\times
exp(-(2-\gamma)\gamma^{\frac{\gamma}{2-\gamma}}(\frac{|x|^{2+\theta}}{(2+\theta)^2
Dt^{\gamma}})^{\frac{1}{2-\gamma}}).
\end{equation}
\begin{center}
\begin{figure}[thb]
\scalebox{1.3}[1.1]{\includegraphics{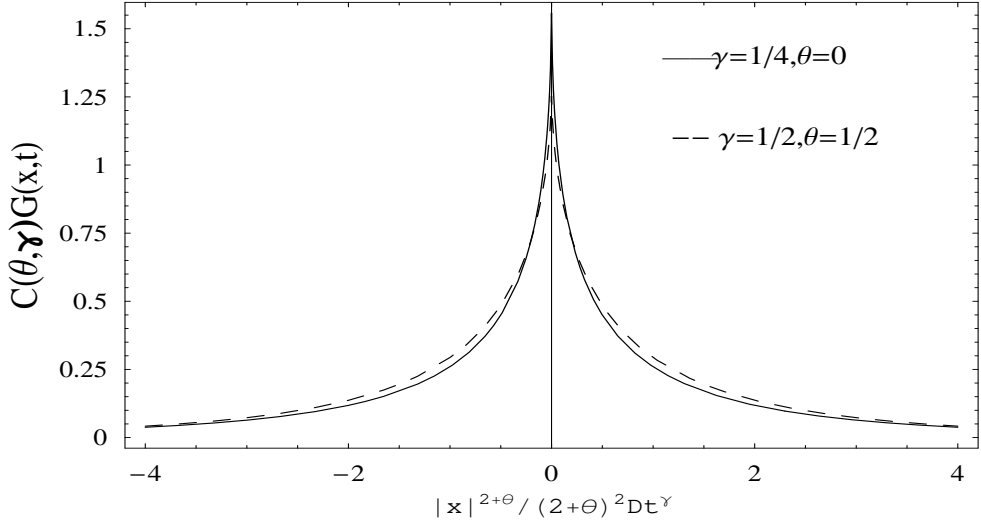}}
 \caption{\label{Fig.1}
\small{The behavior of green function $\mathcal{G}(x,t)$ in Eq.(12)
is illustrated by considering $C(\theta,\gamma)\mathcal{G}(x,t)$
versus $\frac{|x|^{2+\theta}}{(2+\theta)^2 Dt^{\gamma}}$ for typical
values of $\gamma$ and $\theta$. Here
$C(\theta,\gamma)=\frac{2\Gamma(1/(2+\theta))}{2+\theta}((2+\theta)^2
Dt^{\gamma})^{1/(2+\theta)}$.
 }}
 \end{figure}
\end{center}
In this direction, Eq.(14) can be considered as an extension of the
asymptotic behavior of homogeneous and isotropic random walk
models [26].\\
\textbf{  Case 2. } $D(t)=\frac{Dt^{\alpha-1}}{\Gamma(\alpha)}$,
i.e.$\tilde{D}(s)=Ds^{-\alpha}$.

By using the same method as in case 1, we obtain
\begin{equation}
\mathcal{G}(x,t)=\frac{2+\theta}{2\Gamma(\frac{1}{2+\theta})}(\frac{1}{(2+\theta)^2
Dt^{\gamma+\alpha}})^{\frac{1}{2+\theta}}H_{1 \quad 2}^{2\quad
0}[\frac{|x|^{2+\theta}}{(2+\theta)^2
Dt^{\gamma+\alpha}}|_{(0,1),(\frac{1+\theta}{2+\theta},1)}^{(1-\frac{\gamma+\alpha}{2+\theta},\gamma+\alpha)}],
\end{equation}
and
\begin{equation}
\rho(x,t)=\frac{2+\theta}{2\Gamma(\frac{1}{2+\theta})}(\frac{1}{(2+\theta)^2
Dt^{\gamma+\alpha}})^{\frac{1}{2+\theta}}\int_{-\infty}^{+\infty}dx'\tilde{\rho}(x')H_{1
\quad 2}^{2\quad 0}[\frac{|x-x'|^{2+\theta}}{(2+\theta)^2
Dt^{\gamma+\alpha}}|_{(0,1),(\frac{1+\theta}{2+\theta},1)}^{(1-\frac{\gamma+\alpha}{2+\theta},\gamma+\alpha)}].
\end{equation}

Let us go back to Eq.(1), and consider the external force
$F(x)\varpropto x|x|^{\alpha-1}$, $D(x,t)=D\delta(t)|x|^{-\theta}$
and $\mu=2$, $\nu=1$. In this case ,analytical solution can not
easily be obtained for a generic $\alpha$, $\theta$. However, for
$\theta\neq 0$, and $\alpha+\theta+1=0$. By following the same
procedure as in the above case, an exact solution can be obtained
and it is given by
\begin{equation}
\rho(x,t)=\frac{2+\theta}{2\Gamma(\frac{1}{2+\theta}+\frac{\mathcal
{K}}{(2+\theta)D})}(\frac{1}{(2+\theta)^2
Dt^{\gamma}})^{\frac{1}{2+\theta}}H_{1 \quad 2}^{2\quad
0}[\frac{|x|^{2+\theta}}{(2+\theta)^2 Dt^{\gamma}}|_{(\frac{\mathcal
{K}}{(2+\theta)D},1),(\frac{1+\theta}{2+\theta},1)}^{(1-\frac{\gamma}{2+\theta},\gamma)}],
\end{equation}
where, for simplicity, we are considering the initial condition
$\rho(x,0)=\delta(x)$, and the external force (drift)
$F(x)=\mathcal{K}x^{\alpha}$. The second moment is given by
$<x^2>\varpropto t^{\frac{2\gamma}{2+\theta}}$, which scales with
the exponent $\frac{2\gamma}{2+\theta}$ and clearly depends only on
$\gamma$ and $\theta$. So, when $\frac{2\gamma}{2+\theta}<1$, $=1$
and $>1$, the system is sub-diffusion, normal diffusion and
supper-diffusion respectively.

The presence of the external force in Eq.(1) is now changed into
$F(x)=-k_1 x+k_2 x^{-1-\theta}$. In order to obtain the solution of
Eq.(1), we expand $\rho(x,t)$ in terms of the eigenfunctions, i.e.
we employ $\rho(x,t)=\sum_{n=0}^{\infty}\phi_n (t)\psi_n (x)$ with
$\psi_n (x)$ determined by the spatial equation
\begin{equation}
-\lambda_n \psi_n (x)=D\frac{\pa}{\pa x}\{x^{-\theta}\frac{\pa}{\pa
x}\psi_n (x)\}+\frac{\pa}{\pa x}\{(k_1 x+k_2 x^{-1-\theta})\psi_n
(x)\}
\end{equation}
and $\phi_n (t)$ determined by the time equation
\begin{equation}
\frac{\pa^{\gamma}}{\pa t^{\gamma}}\phi_n (t)=-\lambda_n \phi_n (t).
\end{equation}
The solution for the time equation is given by in terms of the
Mittag-Leffter function
\begin{equation}
\phi (t)\propto E_{\gamma}(-\lambda_n t^{\gamma}).
\end{equation}
In order to get the solution for Eq.(18), we perform the transform
\begin{equation}
\psi_n (x)=e^{-y}y^{\delta}Z(y),\quad y=Ax^{v}
\end{equation}
to translate Eq.(18) into the associated Laguerre equation
\begin{equation}
yZ''(y)+(\alpha+1-y)Z'(y)+nZ(y)=0
\end{equation}
with parameter $\alpha$ under the following condition:
\begin{equation}
\alpha=\frac{\frac{k_2}{D}-1-\theta}{2+\theta},v=2+\theta,A=\frac{k_2}{(2+\theta)D},\delta=\frac{k_2}{(2+\theta)D},\lambda_n
=(2+\theta)nk_1.
\end{equation}
Then, using the Green function methods and after some calculations,
it is possible to show that
\begin{equation}
\rho(x,t)=\int_{-\infty}^{\infty}dx_0
\tilde{\rho}(x_0)\mathcal{G}(x,x_0,t),
\end{equation}
$$\mathcal{G}(x,x_0,t)=(\frac{k_1}{(2+\theta)D})^{\frac{k_2
+D}{(2+\theta)D}}|x|^{\frac{k_2}{D}}e^{-\frac{k_1}{(2+\theta)D}|x|^{2+\theta}}\sum_{n=0}^{\infty}\frac{(2+\theta)\Gamma(n+1)}{\Gamma(n+\frac{k_2
+D}{(2+\theta)D})}E_{\gamma}(-\lambda_n t^{\gamma})$$
\begin{equation}
\times L_{n}^{(\alpha)}(\frac{k_1}{(2+\theta)D}|x|^{2+\theta})
L_{n}^{(\alpha)}(\frac{k_1}{(2+\theta)D}|x_0|^{2+\theta}),
\end{equation}
where $L_n^{(\alpha)} (x)$ is the associated Laguerre polynomial and
it is the solution for  Eq.(22). Here, we used the formula
\begin{equation}
\int_0^\infty dy
y^{\alpha}e^{-y}L_n^{(\alpha)}(y)L_n^{(\alpha)}(y)=\frac{\Gamma(n+\alpha+1)}{\Gamma+1}.
\end{equation}
Notice that Eq.(25) contains the usual Ornstein-Uhlenbeck process
[27] and the usual Rayleigh process [28] as particular cases and it
extends the results obtained in [21].  In this context, in the
presence of an constant absorbent force, i.e.
$\tilde{\alpha}\rho(x,t)$, we can obtain the solution for Eq.(1)
only need to change the argument of the Mittag-Leffler function
present in Eq.(20) to $\lambda_n +\tilde{\alpha}$.

Let us now discuss Eq.(1) by considering a mixing between the
spatial and time fractional derivatives. For simplicity, we consider
Eq.(1) in the absence of the external force with
$(\nu,\theta)=(1,0)$ and $\gamma$, $\mu$ arbitraries. Applying the
Fourier and Laplace transform to Eq.(1) and employing the Riez
representation for the spatial fractional derivatives, we have
\begin{equation}
s^{\gamma}\hat{\tilde{\rho}}(k,s)-s^{r-1}\hat{\rho}(k,0)=-\tilde{D}(s)|k|^{\mu}\hat{\tilde{\rho}}(k,s),
\end{equation}
where $\hat{\rho}(k,t)=\mathcal{F}\{\rho(x,t)\}=\int_{-\infty}
^{+\infty} \rho(x,t)e^{-ikx}dx$ , so $\rho(k,0)=1$. Here, we
consider the diffusion coefficient given by
$D(t)=\frac{Dt^{\alpha-1}}{\Gamma(\alpha)}$, i.e.
$\tilde{D}(s)=Ds^{-\alpha}$ and $\rho(x,0)=\delta(x)$. By using the
inverse of  Laplace transform, we obtain
$$\rho(k,t)=E_{\gamma+\alpha, 1}(-D|k|^{\mu}t^{\gamma+\alpha})$$
\begin{equation}
\quad \quad \quad \quad \quad \quad \quad=H_{1 \quad 2}^{1\quad
1}[D|k|^{\mu}t^{\gamma+\alpha}|_{(0,1),(0,\gamma+\alpha)}^{(0,1)}].
\end{equation}
This solution recovers the usual one for $(\mu,\gamma)=(2,1)$ and
for $\mu\neq 2$ it extends the results found in [29]. In order to
perform the inverse of Fourier transform, we employ the procedure
presented in [30]. Then we can obtain
\begin{equation}
\rho(x,t)=\frac{1}{2\mu
\sqrt{\pi}(Dt^{\gamma+\alpha})^{\frac{1}{\mu}}}H_{2 \quad 3}^{2\quad
1}[\frac{|x|}{2(Dt^{\gamma+\alpha})^{\frac{1}{\mu}}}|_{(0,1/2),(1-1/\mu,1/\mu),(1/2,1/2)}^{(1-1/\mu,1/\mu),(1-(\gamma+\alpha)/\mu,(\gamma+\alpha)/\mu)}].
\end{equation}
The stationary solution that emerges from this process is a Levy
distribution.

Now, we consider a particular case of Eq.(1) for $\gamma=1$ and
nonzero values of $\mu$ and $\theta$, and consider a linear
drift,i.e. $F(x)=-\mathcal {K}x$. For simplicity, we employ
$D(t)=D\delta(t)$ and the initial condition $\rho(x,0)=\delta(x)$,
then Eq.(1) yields to
\begin{equation}
\frac{\pa}{\pa t}\rho(x,t)=D\frac{\pa}{\pa
x}\{|x|^{-\theta}\frac{\pa ^{\mu-1}}{\pa
x^{\mu-1}}[\rho(x,t)]^{\nu}\}+\frac{\pa}{\pa x}\{\mathcal
{K}x\rho(x,t)\}.
\end{equation}
Let us investigated time dependent solutions for Eq.(30). We use
similarity methods to reduce Eq.(30) to ordinary differential
equations. The explicit form for these ordinary differential
equations depends on the boundary conditions or restrictions in the
form of conservation laws. In this direction, we restrict our
analysis to find solution that can be expressed as a scaled function
of the type [31]
\begin{equation}
\rho(x,t)=\frac{1}{\phi(t)}\tilde{\rho}(z),\quad
z=\frac{x}{\phi(t)}.
\end{equation}
Inserting Eq.(31) into Eq.(30), we obtain
\begin{equation}
-(\frac{\dot{\phi(t)}}{\phi(t)^2}+\frac{\mathcal{K}}{\phi(t)})\frac{\pa}{\pa
z}[z\tilde{\rho}(z)]=\frac{D}{\phi(t)^{\theta+\mu+\nu}}\frac{\pa}{\pa
z}[z^{-\theta}\frac{\pa^{\mu-1}}{\pa
z^{\mu-1}}\tilde{\rho}(z)^{\nu}].
\end{equation}
By choosing the ansatz
\begin{equation}
\frac{\dot{\phi(t)}}{\phi(t)^2}+\frac{\mathcal{K}}{\phi(t)}=\frac{kD}{\phi(t)^{\theta+\mu+\nu}},
\end{equation}
where $k$ is an arbitrary constant which can be determined by the
normalization condition. By solving Eq.(33), we have that
\begin{equation}
\phi(t)=[(\phi(0))^{\theta+\mu+\nu-1}e^{-(\theta+\mu+\nu-1)\mathcal{K}t}+\frac{Dk}{\mathcal{K}}(1-e^{-(\theta+\mu+\nu-1)\mathcal{K}t})]^{\frac{1}{\theta+\mu+\nu-1}}.
\end{equation}
By substituting Eq.(33) into Eq.(32), we obtain
\begin{equation}
\frac{\pa}{\pa z}[z^{-\theta}\frac{\pa^{\mu-1}}{\pa
z^{\mu-1}}\tilde{\rho}(z)^{\nu}]=-k\frac{\pa}{\pa
z}[z\tilde{\rho}(z)].
\end{equation}
Then, we perform an integration and the result is
\begin{equation}
z^{-\theta}\frac{\pa^{\mu-1}}{\pa
z^{\mu-1}}\tilde{\rho}(z)^{\nu}=-kz\tilde{\rho}(z)+\mathcal {C},
\end{equation}
where $\mathcal{C}$ is another arbitrary constant. Also, we use the
following generic result [32]:
\begin{equation}
D_{x}^{\delta}[x^{\alpha}(a+bx)^{\beta}]=a^{\delta}\frac{\Gamma[\alpha+1]}{\Gamma[\alpha+1-\delta]}x^{\alpha-\delta}(a+bx)^{\beta-\delta}
\end{equation}
with $D_{x}^{\delta}\equiv d^{\delta}/dx^{\delta}$ and
$\delta=\alpha+\beta+1$. By defining $g(x)\equiv
x^{\frac{\alpha}{\nu}}(a+bx)^{\frac{\beta}{\nu}}$ and $\lambda\equiv
\alpha(1-\frac{1}{\nu})-\delta$, and rearranging the indices,
Eq.(37) can be rewritten as follows:
\begin{equation}
D_{x}^{\delta}[g(x)]^{\nu}=a^{\delta}\frac{\Gamma[\alpha+1]}{\Gamma[\alpha+1-\delta]}x^{\lambda}g(x).
\end{equation}
For this case, we consider the ansatz
$\tilde{\rho}(z)=\mathcal{N}z^{\frac{\alpha}{\nu}}(1+bz)^{\frac{\beta}{\nu}}$.
By using the property of Eq.(38) in Eq.(36) and ,for simplicity,
choosing $\mathcal{C}=0$, we find
$$\alpha=\frac{(2-\mu)(\mu+\theta)}{1-2\mu-\theta},$$
\begin{equation}
\beta=-\frac{(\mu-1)(\mu-2)}{1-2\mu-\theta},
\end{equation}
$$\nu=\frac{2-\mu}{1+\mu+\theta}.$$
In this case, we have
\begin{equation}
\rho(x,t)=\frac{\mathcal{N}}{\phi(t)}[\frac{z^{(\mu+\theta)(1+\mu+\theta)}}{(1+bz)^{(1-\mu)(1+\mu+\theta)}}]^{\frac{1}{1-2\mu-\theta}},
\end{equation}
where $\phi(t)$ is given above,
$\mathcal{N}=[-k\frac{\Gamma(-\beta)}{\Gamma(\alpha+1)}]^{\frac{\mu+\theta+1}{1-2\mu-\theta}}$
and $b$ is an arbitrary constant (to be taken, later on, as $\pm1$
according to the specific solutions that are studied). Several
regions can be analyzed. For simplicity, we illustrate two of them:
$-\infty<\mu<-1-\theta$ with $\theta\geq0$, and $0<\mu<1/2$ with
$0\leq\theta<1/2-\mu$. Let us start by considering the region
$-\infty<\mu<-1-\theta$.  Without loss of generality, we choose
$b=-1$. The normalization condition implies(see Fig. 2)
\begin{equation}
\mathcal{N}\int_{-1}^{1}[\frac{z^{(\mu+\theta)(1+\mu+\theta)}}{(1+bz)^{(1-\mu)(1+\mu+\theta)}}]^{\frac{1}{1-2\mu-\theta}}dz=1.
\end{equation}
So
\begin{equation}
\mathcal{N}=\frac{\Gamma[1-\mu-\theta]}{2\Gamma[\frac{\mu^2+\mu\theta-2\theta-2\mu}{1-2\mu-\theta}]\Gamma[\frac{1-\mu+\mu^2+\theta^2+2\mu\theta}{1-2\mu-\theta}]}.
\end{equation}
\begin{center}
\begin{figure}[thb]
\scalebox{1.3}[1.1]{\includegraphics{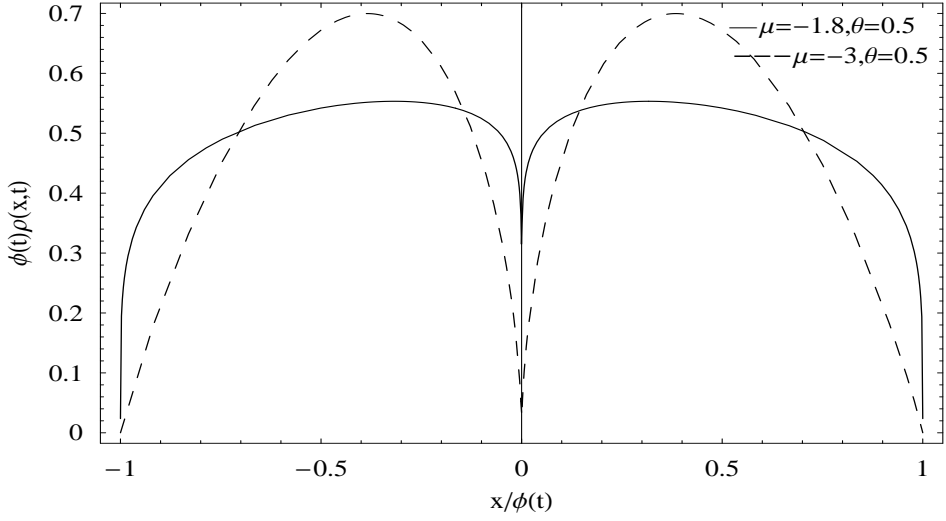}}
 \caption{\label{FigII}
\small{Behavior of $\phi(t)\rho(x,t)$ versus $x/\phi(t)$, which
illustrates Eq.(40) with typical values for $\mu$ and $\theta$
satisfying $0<\mu<-1-\theta$ and $\theta\geq0$. We notice that the
distribution vanishes at the abcissa equal $\pm1$, and remains zero
outside of this interval.
 }}
 \end{figure}
\end{center}
Let us now analyze the region $0<\mu<1/2$ with
$0\leq\theta<1/2-\mu$. Again without the loss of generality, we
choose $b=1$. The normalization condition implies(see Fig. 3)
\begin{equation}
\mathcal{N}=\frac{\Gamma[\frac{1+\theta-\mu^2-\mu\theta}{1-2\mu-\theta}]}{2\Gamma[\frac{1-\mu+\mu^2+\theta^2+2\mu\theta}{1-2\mu-\theta}]\Gamma[\mu+\theta]}.
\end{equation}
\begin{center}
\begin{figure}[thb]
\scalebox{1.3}[1.1]{\includegraphics{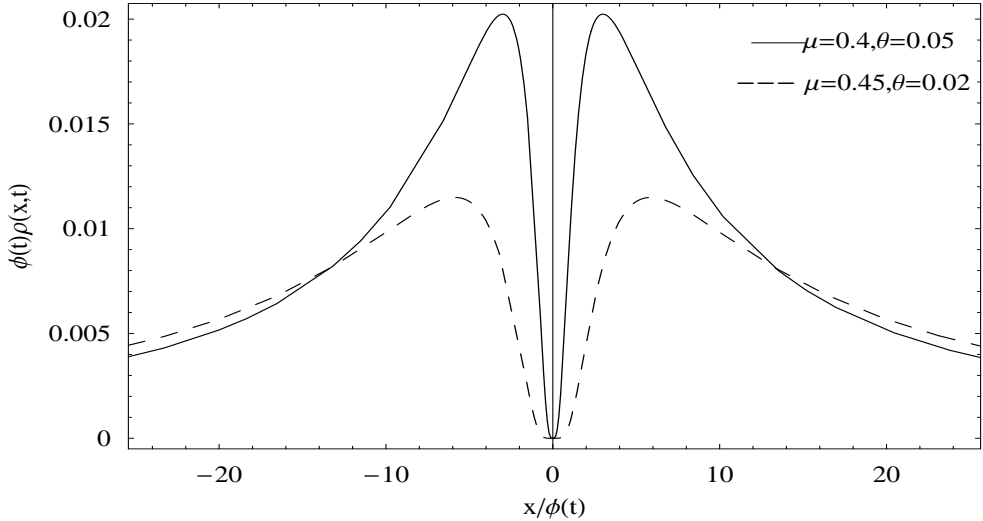}}
 \caption{
\small{Behavior of $\phi(t)\rho(x,t)$ versus $x/\phi(t)$, which
illustrates Eq.(40) with typical values for $\mu$ and $\theta$
satisfying $0<\mu<1/2$ and $0\leq\theta<1/2-\mu$. We notice that the
distribution vanishes at the infinity.
 }}
 \end{figure}
\end{center}

Let us finally mention a connection between the results obtained
here and the solutions that arise from the optimization of the
nonextensive entropy [33]. These distributions do not coincide for
arbitrary value of x. However, the comparison of the
$|x|\rightarrow\infty$ asymptotic behaviors enables us to identify
the type of Tails. By identifying the behavior exhibited in Eq.(40)
with the asymptotic behaviors $1/|x|^{2/(q-1)}$ that appears in [33]
for the entropic problem, we obtain
\begin{equation}
q=\frac{3+\mu+\theta}{1+\mu+\theta}.
\end{equation}
This relation recovers the situation for $\theta=0$ .

\section{Summary and Conclusions}

We have analyzed the generalized fractional diffusion equations by
considering an external force $F(x)\varpropto x|x|^{\alpha-1}$ and a
spatial time-dependent diffusion coefficient
$D(x,t)=D(t)|x|^{-\theta}$. By using Laplace transform, Fourier
transform, the Green function method and normalized scaled function
we can find the explicit solutions $\rho(x,t)$ which subjects to the
natural boundary condition $\rho(\pm\infty,t)=0$ and the initial
condition $\rho(x,t)=\tilde{\rho}(x)$. In a word, we have extended
the results previously obtained by the other authors by taking an
external force and a spatial time-dependent diffusion coefficient
into account. We have also discussed the connection with
nonextensive statistics, providing the relation between our
solutions and those obtained within the maximum entropy principle by
using the Tsallis entropy. Finally, we expect that the results
obtained here may be useful to the discission of the anomalous
diffusion systems where fractional diffusion equations play an
important role.

{\bf Acknowledgments:} The authors are grateful to the anonymous
referees for useful comments and suggestions.


\thebibliography{00}
 \baselineskip 13pt

\bibitem{mur90} M. Muskat, The Flow of Homogeneous Fluid Through Porous Media, McGraw-Hill, New York, 1937.

\bibitem{pol96} P.Y. Polubarinova-Kochina, Theory of Ground Water Movement, Princeton University Press,Princeton, 1962.

\bibitem{gro06} P. Grosfils and J.P. Boon, Physica A 362 (2006) 168.

\bibitem{buc77} J. Buckmaster, J.Fluid Mech. 81 ( 1977) 735.

\bibitem{plo04} S.S. Plotkin and P.G. Wolynes, Phys. Rev. Lett. 80 (1998) 5015 .

\bibitem{cro04} D.S.F. Crothers, D. Holland, Y.P. Kalmykov and W.T. Coffey, J. Mol. Liq. 114 (2004) 27.

\bibitem{mez99} R. Metzler, E. Barkai and J. Klafter, Physica A 266 (1999) 343.

\bibitem{cam04} D. Campos, V. Mendez and J. Fort, Phys. Rev. E 69
(2004) 031115.

\bibitem{wan07} M. Wang and S.Y. Chen, J. Colloids Interface Sci. 314
(2007) 264.

\bibitem{wan06} M. Wang, JK. Wang, S.Y. Chen and N. Pan, J. Colloids Interface
Sci. 304 (2006) 246.

\bibitem{bor98} L. Borland, Phys. Rev. E 57 (1998) 6634.

\bibitem{shl94} M.F. Shlesinger, G.M. Zaslavsky and U. Frisch, L\'{e}vy Flights and Related Topics in Physics,
Springer-Verlag, Berlin, 1994.

\bibitem{sch89} W.R. Schneider and W. Wyss, J. Math.
Phys. 30 (1989) 134.

\bibitem{met00} R. Metzler and J. Klafter, Physica A 278 (2000) 107.

\bibitem{mai03} F. Mainard and G. Pagnini, Appl. Math.
Comput. 141 (2003) 51.

\bibitem{dra00}G. Drazer, H.S. Wio and C. Tsallis, Phys. Rev.
E 61 (2000) 1417.

\bibitem{len05} E.K. Lenzi, R.S. Mendes, J.S. Andradi, L.R. da Silva
and L.S. Lucena, Phys. Rev. E 71 (2005) 052109.

\bibitem{pod99} I. Podlubny, Fractional differential equations, Academic
Press, San Diego, CA, 1999. 54

\bibitem{lez03} E.K. Lenzi, L.C. Malacarne, R.S. Mendes and I.T.
Pedron, Physica A 319 (2003) 245.

\bibitem{spo93} H. Spohn, J. Phys. 13 (1993) 69.

\bibitem{len03} E.K. Lenzi, R.S. Mendes, Kwok Sau Fa and
L.C.Malacame, J. Math. Phys. 44 (2003) 2179.

\bibitem{len04} E.K. Lenzi, R.S. Mendes and Kwok Sau Fa, J. Math. Phys.
45 (2004) 3444.

\bibitem{mor53} M.P. Morse and H. Feshbach, Methods of Theoretical Physics,
McGraw-Hill, New York, 1953.

\bibitem{ren06} F.Y. Ren, J.R. Liang, W.Y. Qiu and J.B. Xiao,
J.Phys.A:Math.Gen, 39 (2006) 4911.

\bibitem{mat78} A.M. Mathai and R.K. Saxtena, The H-function with
Application in Statistics and Other Disciplines, Wiley Eastern, New
Delhi, 1978.

\bibitem{kla91} J. Klafter, G. Zumoften and A. Blumen, J.phys.A 25 (1991)
4835.

\bibitem{uhl30} G.E. Uhlenbeck and L.S. Ornstein, Phys.Rev.E 36 (1930)
823.

\bibitem{gar96} C.W. Gardiner, Handbook of Stochastic Methods: for
Physics, Chemistry and the natural Sciences, Springer Series in
Synergetics , Springer, New York, 1996.

\bibitem{met99} R. Metzler, E. Barkai and J. Klafter, Phys. Rev. Lett.
82 (1999) 3563.

\bibitem{lan06} T.A.M. Langlands, Physica A 367 (2006) 136.

\bibitem{tsa02} C. Tsallis and E.K. Lenzi, Chem. phys
. 284 (2002) 341.

\bibitem{bol00} M. Bologna, C. Tsallis and P. Grigolini, Phys.Rev.E
62 (2000) 2213.

\bibitem{tsa96} C. Tsallis and D.J. Bukman, Phys.Rev.E
54 (1996) R2197.
\end{document}